\documentclass{sig-alt-release2}

\usepackage{hyperref}

\begin{document}

\conferenceinfo{WWW 2013 Companion,} {May 13--17, 2013, Rio de Janeiro, Brazil.} 
\CopyrightYear{2013} 
\crdata{978-1-4503-2038-2/13/05} 
\clubpenalty=10000 
\widowpenalty = 10000

\title{ResourceSync: Leveraging Sitemaps for Resource Synchronization}
%
%
%
%
%

\numberofauthors{6} 
%
\author{
%
%
\alignauthor Bernhard Haslhofer\\
\affaddr{Cornell University, Information Science}\\
\email{bh392@cornell.edu}
\alignauthor Simeon Warner\\
\affaddr{Cornell University Library}\\
\email{simeon.warner@cornell.edu}
\alignauthor Carl Lagoze\\
\affaddr{University of Michigan, School of Information}\\
\email{clagoze@umich.edu}
\and
\alignauthor Martin Klein\\
\affaddr{Los Alamos National Laboratory}\\
\email{mklein@lanl.gov}
\alignauthor Robert Sanderson\\
\affaddr{Los Alamos National Laboratory}\\
\email{rsanderson@lanl.gov}
\alignauthor Michael L. Nelson\\
\affaddr{Old Dominion University}\\
\email{mln@cs.odu.edu}
\and
\alignauthor Herbert Van de Sompel\\
\affaddr{Los Alamos National Laboratory}\\
\email{herbertv@lanl.gov}
}

\maketitle
\begin{abstract}

Many applications need up-to-date copies of collections of changing Web resources. Such synchronization is currently achieved using ad-hoc or proprietary solutions. We propose ResourceSync, a general Web resource synchronization protocol that leverages XML Sitemaps. It provides a set of capabilities that can be combined in a modular manner to meet local or community requirements. We report on work to implement this protocol for arXiv.org and also provide an experimental prototype for the English Wikipedia as well as a client API.

\end{abstract}



\section{Introduction}

Synchronization of resources from one Web-based system, a \emph{source}, to another, a \emph{destination}, is frequently important. It may be necessary to ensure reliable access to a set of resources, to provide backup copies for preservation purposes, or to leverage computational resources or tools available at one server but not another. We consider three examples of synchronization used by popular services:

\begin{itemize}
    \item The arXiv.org collection of physics, mathematics, and computer science articles that exist on a primary server but are mirrored at other servers worldwide.
    \item Structured Web data sources such as DBPedia that are synchronized with changes in their unstructured counterparts (Wikipedia).
    \item The data.europeana.eu service, which aggregates metadata from many remote sources across Europe.
\end{itemize}

The synchronization problem is well-known and various solutions are available. One can use \emph{rsync}~\cite{tridgell1996rsync} to synchronize local and remote file systems, use OAI-PMH\footnote{\url{http://www.openarchives.org/pmh/}} to synchronize metadata between repositories, synchronize files between two machines via WebDAV\footnote{\url{http://tools.ietf.org/html/rfc4918}}, or install Dropbox or Google drive to synchronize local files or documents with cloud services. However, all of these mechanisms are proprietary or ad-hoc solutions, which position the important function of synchronization as an outlier among other functionality that has been comfortably incorporated with the Web Architecture (e.g., REST and Linked Data).

We propose ResourceSync~\cite{van2012perspective,klein2013technical} as a framework for a Web-based synchronization mechanism, which leverages the wide-spread adoption of XML Sitemaps\footnote{\url{http://www.sitemaps.org/}}. It provides a modular set of synchronization components for baseline synchronization, incremental synchronization, and pull- or push-based change awareness that can flexibly be combined to meet a variety of synchronization requirements. Our presentation will focus on the following issues: 

\begin{enumerate}

    \item We introduce the ResourceSync framework, which is now available in a first beta draft (\texttt{http://www.open\-archives.org/rs/0.5/}). In the simplest case, it requires only that a source exposes an XML Sitemap listing resources with last modification information.
    
    \item We report on the experiences from implementing ResourceSync for arXiv.org and the English Wikipedia and discuss how it might support information aggregation from many diverse sources in Europeana.
    
\end{enumerate}

All ResourceSync implementations, including a Python client library and a simulator for testing purposes, are available in a Git repository: \url{https://github.com/resync}.

\section{Synchronization Scenarios}\label{sec:use_cases}

\subsection{arXiv.org}

arXiv (\url{http://arXiv.org})\footnote{The arXiv statistics come from Simeon Warner, one of the authors of this paper.} is a well-known and heavily used repository of scholarly articles in physics, mathematics, computer science and related disciplines. It has over 800,000 articles with an average of about 1.5~revisions per article. For each version there is a separate metadata record and the full-text package (PDF, tar.gz, etc.) giving a total of approximately 2.4~million resources. New articles and revisions are added at the rate of about 75,000 per year, and there are also occasional metadata changes such as the addition of bibliographic information for journal versions of articles. Updates are made public at 8pm eastern US time each day Sunday through Thursday, with an average of 1,800 resources updated at that moment at that time.

We consider two synchronization use cases: the first is the synchronization of content to mirror sites\footnote{\url{http://arXiv.org/help/mirrors}} which are under direct arXiv control, and the second is synchronization of content to other independent services\footnote{Example services include the UK Institute of Physics \url{http://eprintWeb.org/} site, or the Math Front at UC Davis \url{http://front.math.ucdavis.edu/}} or to researchers for bibliometric and scientometric analysis. In the first case, the goals are high consistency, moderate latency, and robustness to global network outages. There is also the need for the system to automatically recover from outages without human intervention. In the second case, the goal is to make resource and update information publicly available so that any other service may synchronize at the frequency it needs without the need for any out-of-band communication.

The current mirroring system 
uses a process of an HTTP trigger from the main site, an HTTP pull of a list of changed objects specific to the particular mirror site, HTTP download of the resources listed, an HTTP request to verify when the mirror has completed downloading, and then verification (via HTTP HEAD) by the main site which updates the list of out-of-sync items for the particular mirror. The process is periodically repeated as long as there are updates for that mirror. This system is limited to a trusted set of servers operating with the same internal organization and is not available publicly. It also does not support an audit process to check synchronization so rsync is used periodically. Switching to a standardized resource-centric framework could combine efficient updates, the ability to do periodic audits, public synchronization capability, and reduce the burden of maintaining an ad-hoc system.

\subsection{wikipedia.org}

The multilingual online encyclopedia Wikipedia currently contains over 23~million articles in various languages and is maintained by about 100,000 active contributors. There are 285 language editions: the English Wikipedia with 4 million articles is the largest one, followed by the German, French, and Dutch editions, which all have between 1~and 4~million articles\footnote{\url{http://en.wikipedia.org/wiki/Wikipedia}}. According to measurements reported in~\cite{hellmann2009dbpedia}, about 1.4 article pages are updated each second on Wikipedia which amounts to 120,000 page updates per day.

We envision two main synchronization use cases for Wiki\-pedia: first there are large structured knowledge bases such as Freebase and DBPedia which are increasingly used in combination with information retrieval techniques (e.g., DBPedia Spot\-light~\cite{mendes2011dbpedia}, Google Knowledge Graph\footnote{\url{http://www.google.com/insidesearch/features/search/knowledge.html}}). They heavily reuse or even mirror information from Wikipedia and therefore need to refresh this information when remote resources change. Second, there are publishers and media agencies like the New York Times or the BBC, which link entities in their information space with entities in Wikipedia and reuse information (e.g., article abstracts) in their own information space. Keeping these information sources in sync is an important component of sustaining the timeliness of the news source.

Currently Wikipedia exposes article metadata and change information via a non-public OAI-PMH endpoint, and some editions also push article change information via a dedicated IRC channel. This means that there is no uniform, Web-based solution that allows clients to replicate and periodically synchronize information from Wikipedia.

\subsection{data.europeana.eu}

Europeana provides access to more than 20~million books, paintings, films, museum objects and archival records that have been digitized throughout Europe, gathered from hundreds of individual institutions, with the help of dozens of data aggregators and providers. The initial Linked Open Data release contains metadata on approximately 2.4~million texts, images, videos and sounds. These collections encompass more than 200~cultural institutions from 15~countries. While the 10~largest data providers contribute 80\% of all data, the remaining 20\% are contributed by smaller institutions. Two data providers even contribute only one single object to the current dataset~\cite{haslhofer2011data}.

In Europeana, Web-based resource synchronization can serve two purposes: first, Europeana internally needs to periodically aggregate metadata from its data providers. Since Europeana provides an entry point for search and retrieval over aggregated objects, it is in the data provider's interest that Europeana has a consistent view over these objects. Second, Europeana could also provide a synchronization endpoint for external services that consume and make use of data provided by Europeana.

At the moment, the Europeana-internal metadata aggregation mechanism is based on OAI-PMH and manual data transfer from the data providers to Europeana, possibly via intermediate aggregators. External data consumers can download data or use available Europeana data dumps but have no means to synchronize resources.

\section{ResourceSync}

In order to allow a destination to initially synchronize with a source, the synchronization process at the destination must be able to retrieve a list of source resources for which synchronization is intended --- we denote this process as \emph{baseline synchronization}. Subsequently it can perform \emph{incremental synchronization} to keep its copies in-sync with the corresponding resources at the source. Finally, the destination can perform an \emph{audit} to check that its copies match the corresponding resources at the source. The central framework components enabling these processes are \emph{resource lists} and \emph{change lists}. ResourceSync also supports the notion of dumps for packaging resources, cross-linkage of related resources, patching resource representations, etc. Further information on these and other capabilities is given in the ResourceSync beta draft.

\subsection{Resource List}

Baseline synchronization requires that a source exposes the list of resource URIs it conveys for synchronization. In its most basic form, as shown in the following listing, a resource list is an XML Sitemap with an additional element expressing that the Sitemap implements resource list capability. Since a resource list presents a snapshot of a Source's resources at a particular point in time, this element also carries the datetime of the resource list's most recent update.

\begin{verbatim}
<?xml version="1.0" encoding="UTF-8"?>
<urlset xmlns="..."
        xmlns:rs="...">
  <rs:md capability="resourcelist"
         modified="2013-01-03T09:00:00Z"/>
  <url>
      <loc>http://example.com/res1</loc>
  </url>
  <url>
      <loc>http://example.com/res2</loc>
  </url>
</urlset>
\end{verbatim}

\subsection{Change List}

Incremental synchronization is an optimization over baseline synchronization. If supported by both source and destination, it can reduce latency caused by the transfer of possibly large resource lists. Instead of retrieving the list of available resources, destinations can retrieve atomic resource state change information bundled in change lists. Change lists are also expressed as Sitemaps.

\begin{verbatim}
<?xml version="1.0" encoding="UTF-8"?>
<urlset xmlns="..."
        xmlns:rs="...">
  <rs:md capability="changelist"
         modified="2013-01-03T11:00:00Z"/>
  <url>
      <loc>http://example.com/res2.pdf</loc>
      <lastmod>2013-01-02T13:00:00Z</lastmod>
      <rs:md change="updated"/>
  </url>
  <url>
      <loc>http://example.com/res3.tiff</loc>
      <lastmod>2013-01-02T18:00:00Z</lastmod>
      <rs:md change="deleted"/>
  </url>
</urlset>
\end{verbatim}

Both resource lists and change lists can carry additional meta-information (e.g., hash, content length, mime-type) to facilitate resource synchronization and audit.

\section{Demos}

We have implemented ResourceSync prototypes for two use cases: arXiv.org\footnote{\url{http://arxiv.resourcesync.net}} and the English Wikipedia\footnote{\url{http://en.wikipedia.resourcesync.net}}. Each prototype exposes a resource list and a change list containing information about resource changes. Resource lists are written by a periodic process, which is triggered daily in the case of arXiv and by the availability of a new dump in the case of Wikipedia. The arXiv prototype writes one change list file per day, whereas the Wikipedia prototype keeps the most recent changes in memory and periodically writes them to a file. In the case of Wikipedia changes are recorded from a dedicated IRC channel (\#en.wikipedia), in the case of arXiv the changes are visible in the arXiv database and periodically written out by a batch process. We are also investigating the possibility of implementing ResourceSync for Europeana.

\section{Summary}

We propose ResourceSync, as a general Web resource synchronization protocol that leverages XML Sitemaps. It provides a set of capabilities that can be combined in a modular manner to meet local or community requirements. We are implementing this protocol for arXiv.org and also provide an experimental prototype for the English Wikipedia as well as a client API.

\section{Acknowledgments}

This work is supported through a generous grant from the Alfred P. Sloan Foundation, JISC, as well as a Marie Curie International Outgoing Fellowship (PIOF-GA-2009-252206) within the 7th European Community Framework Programme. We would like to thank Peter Kalchgruber for developing the ResourceSync Wikipedia prototype.


%
\bibliographystyle{abbrv}
\bibliography{citations}  
%
%
\end{document}